# Students' perspectives on computational challenges in physics class

Patti C. Hamerski,[1] Daryl McPadden,[1] Marcos D. Caballero,[1,2] and Paul W. Irving[1]

[1]*Department of Physics and Astronomy, Michigan State University, East Lansing, Michigan 48824, USA*
[2]*Department of Physics and Center for Computing in Science Education,
University of Oslo, N-0316 Oslo, Norway*

High school science classrooms across the United States are answering calls to make computation a part of science learning. The problem is that there is little known about the barriers to learning that computation might bring to a science classroom or about how to help students overcome these challenges. This case study explores these challenges from the perspectives of students in a high school physics classroom with a newly revamped, computation-integrated curriculum. Focusing mainly on interviews to center the perspectives of students, we found that computation is a double-edged sword: It can make science learning more authentic for students who are familiar with it, but it can also generate frustration and an aversion towards physics for students who are not.

## I. INTRODUCTION AND BACKGROUND

There are increasing and widespread pushes to introduce computation to high school students [1–3]. Integrating computational practices with science, technology, engineering, and mathematics (STEM) classrooms gives learners a more realistic view of what it means to do science, and better prepares students for pursuing careers in a world where computation is ubiquitous [4]. These pushes are also associated with changing standards [5] to teach our high school students how to "think computationally" [6]. As the push for integrating computation into classrooms becomes more prevalent, we must reckon with the problem that little is known about how students will take to computation-integrated science. This research study contributes to the effort to find out more about the student perspective towards computation when it is integrated into the science classroom. Here, we focus on a case of students experiencing computational integration in their high school physics class. By detailing what challenges and perspectives students face in this context, we can start to identify how to make computation-integrated K–12 physics more equitable, enjoyable, and beneficial to learning.

For our purposes, we view computational integration as the act of altering the curriculum of a STEM course to incorporate computational modeling, specifically as a tool to learn the STEM subject. In this way, students don't learn to program separately from learning science, but rather they learn science in a new way, through computational modeling. This is a practice that STEM professionals are intimately familiar with [7]; thus, integrating computation makes STEM classes more authentic to future STEM careers. Authenticity is important in the sense that computation provides a way for disciplinary science practices to be featured and learned in the classroom [8,9].

Computational modeling can be integrated in a variety of ways at the K–12 level. For instance, at the high school level, teachers have created models for planetary motion in an attempt to help students make predictions and discover Newton's law of gravitation through experimentation on the model [7]. This approach involved the teacher creating the computational model and the students interacting with it. This integration focused on the practice of using computational models to explore physical phenomena. Separately, a middle school chose to integrate computation into science classes for fourth, fifth, and sixth graders [10]. The students used SCRATCH programming [11] to create simple models of situations of their choice. For example, one student modeled a projectile launched from a seesaw and got real-time feedback from the computer as they constructed the model. Because SCRATCH uses code blocks rather than text, it was easier for students to interpret errors and connect their computational choices to the model they made. Another example of computational integration, at the college level, involved curricular transformation in an introductory undergraduate lab-based course [12]. The labs in this course were redesigned to include one part traditional lab with hands-on equipment, and one part computational modeling with VPYTHON [13]. The integration also included reflection questions to help students make connections between the programming and the open-ended, hands-on experimentation. One benefit to the students was that by learning the fundamentals of VPYTHON, they were able to better visualize the relevant physics concepts in the lab course [12].

Despite the increasingly widespread adoption, what we know about how students learn in computation-integrated settings lags behind the speed of the changing curricula. As stated in a recent report on the state of interdisciplinary computation-integration-based education, "We still know very little about students' thinking and learning as it unfolds with the use of computational tools. At the very least, new tools for thinking and making sense of data call for curriculum resources that consider students' developing computational literacy. With the introduction of this new competency, novel effects might emerge concerning student engagement, motivation, and identity in computationally enhanced classrooms" [7] (p. 9). Essentially, Caballero et al. call for researchers to develop an understanding of how computation impacts the experiences of students, from the perspectives of students.

To date, there has been no in-depth qualitative research on the affective experiences of students in computation-integrated STEM contexts in which to situate our study. We therefore looked to similar work in other contexts. To start, studies on affect and investigations of students' perspectives have been a major focus in the last 30 years in broader STEM education research [14–33]. In particular, previous research in math has examined the affective impact on students when they engage in specific types of activities such as problem solving [14–17]. An example of this is a case study on affective responses during problem solving in a middle school math context [17]. Hannula demonstrated discipline-specific connections between affect and student success, thereby suggesting that attending to student affect in pedagogy offers a way to improve teaching and learning. In the discipline of chemistry education, multiple studies have been carried out that examine student affect or constructs related to it, like self-efficacy [18–21]. In one study on student affect in an undergraduate chemistry lab [21], the authors observed lab classes and asked students about their affective experiences. Galloway et al.'s [21] findings and implications centered around students having complex, multifaceted affective responses. The authors offered several suggestions for teachers to cultivate positive affect and imbue meaning into the oft-rote manner of chemistry lab teaching. This study is important in that it was the first to study affective experiences in chemistry labs with an in-depth, qualitative approach, and the implications had the potential to make a significant impact on student-centered chemistry lab teaching. In particular, the authors drew from Bretz [22] to demonstrate that affect-focused research can provide insight into what students view as "meaningful learning"—an enterprise that combines learning with relevance and represents part of students' motivation to maintain effort in school settings.

Similarly, in physics education, research abounds on students' affective experience, beliefs, and perspectives [23–28,34]. One study points specifically to a gap we are trying in part to address—Gupta et al. [23] argued that there has been a lack of research in physics education on the role of affect in modeling student learning, especially on fine-grain interactions. They made the case that most research on student-centered physics learning focuses on the content they know rather than their feelings about what they are experiencing [23]. To explore what role affect can play in learning, Alsop and Watts [24] looked at how students approached a physics topic (radiation and radioactivity) according to their attitude and perception towards it. Their study found that it was possible to balance "impassioned knowledge and informed feeling" in the learning of physics, which keeps students engaged but not off track. Some affect-based strategies for how to achieve this balance of engagement and learning were explored by Häussler and Hoffman [25] and Erinosho [26], who showed the importance (according to student perspectives) of linking physics with nontraditional and/or out-of-classroom situations [25], providing materials that had concrete, relevant examples [25,26], and working on physics problems where students could collaborate with peers [26]. This set of affect-based, student-centered physics education studies demonstrates the relevance of affect to the field of physics education research, the need for deeper affect-based work [23], and relevance of affect for exploring student perspectives.

Additionally, there have been a number of studies that center on students' experiences in their computer science classes. Gomes and Mendes [35] suggested that students struggle in computer science because the necessary problem-solving strategies are new to students, especially in lower-level undergraduate courses, where a lot of students have their first exposure to computation. On top of that, students in these introductory courses are often experiencing the psychological stress of their first year in college in tandem with developing new ways of problem solving and thinking. From a broader perspective on computation, a study by Jenkins [36] highlighted specific barriers associated with the computational tasks themselves. He described computational difficulties in terms of a set of skills: coding (syntax, semantics, structure, and style), algorithms, and recipes for translating ideas into code. He argued that the hardest part is the novelty of computation; compared to other subjects, students need much more precision to achieve meaningful progress. This requires mastery over coding skills and some degree of expertise with translating ideas into code, both of which are hard to build when it is so easy to write imperfect code, to which the computer provides convoluted feedback or outright rejects.

Much of the research on students' experiences with computation, like the studies from Gomes and Mendes and Jenkins, focuses on the challenges that students face rather than their reactions to and perspectives on those challenges. Bosse and Gerosa [29] built a compilation of research studies centered around learning difficulties in programming settings. Most of the results from their literature review indicated students tend to be worried about learning syntax, variables, error messages, and code comprehension. Students also generally experienced nervousness with unknown coding concepts like functions and parameters, often resulting in students erecting affective barriers against such challenges. For example, when a student realized their code contained a semantic error, they were more likely to give up and not finish the programming activity because semantic errors take a lot of time and effort to identify and fix [29].

In the last decade, computational education research has begun to explore the relationship between affect and the challenges that students face in computer science courses. A relevant literature review focused on qualitative research in computation education [37]; they identified self-efficacy as a useful construct to examine students' experiences in these contexts. However, much of the existing qualitative literature on students' affective responses exists in advanced, undergraduate course contexts rather than more introductory levels. Additionally, they noticed that much of the qualitative work was trying to develop theories about how learning happens in computational settings rather than explore and explain computational difficulties from the perspectives of students. According to this review, there is a need in computation education to research on how students interpret their learning, especially at the introductory and/or K–12 levels [37].

A handful of studies address similar needs, though they are in short supply. Lishinski *et al.* [30] studied students' affective responses to computational challenges and how difficulties can elicit self-efficacy judgments resulting in maladaptive learning strategies. They emphasize the importance of attending to affect in programming environments, writing, "Emotional reactions contribute to a feedback loop process in learning to program, and previous performance impacts future performance both by virtue of the effect that past experiences have on learning, but also via the effect that past experiences have on emotions" [30] (p. 8). A study from Kinnunen and Simon [31] similarly found that students made assessments of their own self-efficacy throughout the duration of computational tasks. Further, they found that affective experiences were the primary feature of computational work that students remembered after class was over. This brought an urgency to studying affect-based challenges in programming contexts.

The following year, Kinnunen and Simon [32] studied in more detail how students' affective responses were tied to their self-efficacy judgments. They found that self-efficacy was determined early in the course when students had their initial failures or successes with computation. They recommended that instructors should deliberately ensure that initial experiences with computation should include several successes because it is so easy to "fail" by writing imperfect code if you don't know how to interpret feedback from the computer, which is often inadvertently masked by confusing error messages. The same authors further studied the disconnect between affective responses and self-efficacy with longitudinal interviews [33]. In their findings they attributed the disconnect to a lack of reflective activities built into the course. They added to their previous recommendations by suggesting that initial computational experiences should incorporate feedback on the entire experience, not just the correctness of the result.

Studies like those from Kinnunen and Simon [31–33] and the recommendations that sprang from them demonstrate the importance of exploring student affect in a given type of learning environment. Eckerdal *et al.* [38] theorized about why computer science learning elicits in students the affective responses that it does. They framed the initial experiences (where students form their self-efficacy beliefs for the first time [32,33]) as comprising a "liminal space." In everyday terms, they asked, how do computer science students cross the threshold to learning? If it takes some persistence and confusion before students find their bearings in a computer science course, what is helping them get over the hump? The authors examined affect and found that as students crossed over the threshold, their feelings about learning computation transformed from hate and fear to euphoria. This implies that teachers can take clues from affect about where students are in the learning process, and even tailor instruction to help them cross the threshold to learning.

While there has been significant research into student affect and experiences in STEM courses, including computer science, this research has traditionally been siloed into separate disciplines. As computation becomes integrated into STEM courses [7,10,12,39–43], it is important to understand the effects of this integration. Recently, there has been some work that addresses the challenges associated with computation-integrated STEM, though not from a student-centered perspective. For example, one study investigated the ways that computational activities could be difficult in a middle school context [44]. The authors justified doing this in a computation-integrated STEM setting, writing, "learning a domain-general programming language and then using it for domain-specific scientific modeling involves a significant pedagogical challenge." They found that certain features, such as the problem-solving process and the syntactic complexity of programming languages, can be leveraged for learning by eliciting reflection on work or alleviated by employing a simpler programming language like Python. Overall, they relied on identifying challenges through observation of computational activities rather than through the perspectives or affective responses of students. The same was true in a study by Vieira *et al.* [45], where the authors evaluated a computation-integrated materials science and engineering course. They found that it can be helpful to integrate computation with student-facing challenges in mind. For example, early in the curriculum students performed poorly on framing and recognizing computational problems, which could be addressed by providing extra scaffolding for problem solving at the start of the course. This study, like Basu et al. [44], based their investigation on performance metrics and features of the computational activities that could be construed as difficult rather than centering student perspectives or affect.

Several more studies in computation-integrated physics took up non-student-centered approaches but did allude to students' experiences at some point in their research processes. Weber and Wilhelm [46] reviewed broadly the history of computational modeling in physics education, and they identified several implementation-based hurdles, such as having students invest significant time to familiarize themselves with the software. This is especially a hurdle in high school settings, where there might not be time to learn a new programming language within an existing curriculum and learning to program could be harder at that level. Leary et al. [47] focused on implementation-based challenges from the perspectives of university faculty. They found several faculty-perceived challenges: students being resistant to learning a new clunky tool, instructors not being able to devote enough time for students to get used to a programming language, instructors not having support from the department, instructors not being able to cover as much content, and instructors not having time to prepare for the new material. The authors relayed from their participants that it was hard as an instructor to prepare for computation because you must learn a lot about the programming language, and it can be hard to make sure it will be accessible to students who have not used it before.

Other studies highlighted the challenges *and* benefits to students of integrating computation into a physics setting. Svensson et al. [48] viewed computation as a type of social semiotic, meaning it can be used to describe many different phenomena and it can produce many different answers to many different questions. In their view, becoming skilled at computation is like learning to communicate with a new language. An example of this is when students comprehend how a line of code that updates position is connected to the physical relationship between velocity and position. The authors argued the challenge lay in students having limited use of computation: even if students are aware of computation's affordances, they might not be able to use computational resources skillfully. On the other hand, with proper guidance or computational experience, students can explore questions and build semiotic resources with code (e.g., conceptual connections and syntactic understanding), and those resources can launch further inquiries. In the authors' view, we need to equip students to see the "affordances" of computational integration. We see a worrying alternative, which is that without an understanding of computation's benefit, students could adopt the view that they have an inability to learn languages (like having a "fixed" mindset [49]), and this could prevent them engaging with computation.

There are additional studies that highlight the student-perceived benefits that computation can bring to STEM classrooms. In an investigation on the impact of a Python-based, university-level computational integration [50], the authors reported that students were excited about learning computation, though the integration didn't have a significant benefit to learning until the second year of physics, when students who had learned the computational tools were able to leverage their proficiency with certain lab tools and data analysis techniques. Caballero et al. [41] highlighted several other benefits that computation brings to physics. They focused their work on high school settings where Modeling Instruction [51] was in use, and they argued that computation highlights relationships between physics concepts, creates dynamic visual models, and can be used to explore real-world, complex physics problems because of its computing power. Furthermore, they explained that students who use computation are learning to use the tools that professional scientists use, which makes physics learning more authentic.

Furthermore, Caballero [52] interviewed professional physicists and physics graduates about how they use computation in everyday work, in an effort to paint a picture of what students should be taught in a computation-integrated physics course. The relevant skills (based on the interviews) were conceptual understanding of physics, writing pseudocode, computational thinking, connecting ideas between math, physics, and computation, understanding the purpose of using computation beyond analytic problems, and learning professional programming practices like writing comments in your code. The interviewees in this study were self-taught programmers, which further shows there is a need for these types of skills to be introduced into physics curricula.

Caballero et al. [7] summarized the research on computation-integrated STEM classes and provided several recommendations for future research and implementations. They argued for the need to (i) develop approachable computational models that reflect modern science so that students can do science using the computation tools, (ii) *study how computation changes student attitudes* and problem solving, (iii) promote proven learning standards when implementing computational integration, and (iv) support teachers as developers of their own content and members of a computation-integrating community.

Thus, we see that research into computation-integrated STEM classes has begun to address the challenges of integration and the impacts on students; however, to our knowledge, there has not been a study that focuses on students' perceptions of the integration and the impacts on their affect, despite its importance in other areas of STEM and multiple calls for research. We intend for this study to begin to fill this gap and to focus specifically on the students' perceptions, challenges, and experiences in a computation-integrated physics course. With this setup in mind, we orient our research question: *What student-perceived, affect-based challenges do high schoolers face in computation-integrated physics?*

In Sec. II, we describe the methodology that drives our use of the analytic tool and our choice around research design which is followed by a description of the study context in Sec. III, including the teacher's choices around computational integration. In Sec. IV we describe our methods for generating data, creating transcripts, and doing analysis. In Sec. V we outline and describe our results, specifically around student-perceived challenges, and we connect our results to affective literature in Sec. VI. In Sec. VII we outline some of the student-perceived benefits of computation, and in Sec. VIII, we discuss our findings and implications of our research. Finally, we conclude in Sec. IX.

## II. METHODOLOGY

Our focus on student perspectives motivates us to use an interpretivist case study lens in this research. We describe this work as a case study because of our variation in data sources and because we aim to capture computational experiences of students in their natural classroom setting. In particular, we take an interpretivist lens because of our focus on students and their perspectives. What defines interpretivist case study [53–55] is how the participants use language. Language constructs the case and shows how the relevant participants interact with the case. Everything that the participants view as meaningful *is* meaningful [54].

The interpretivist approach [54] lends itself well to studies that focus on how people experience and interpret a phenomenon, as opposed to the phenomenon itself. Because we are aiming to open an exploration of *how* students experience computation in their physics class, an interpretivist case study is ideal for exploring this in an in-depth, qualitative way. Using interpretivist case study, we would describe the crux of this study as "how students perceive and react to" affect-based challenges in computation-integrated high school physics with the case being a single physics class taught by Mr. Buford (pseudonym).

In determining our data sources, we bounded the "reality" of our case to the students themselves and classroom occurrences [53,56–58]. For example, we did not study the home life of any students to see how they dealt with their physics obligations outside the classroom. The reason for this bounding was to privilege data sources closest to the phenomenon: student interviews and classroom observations. Though students occasionally mentioned out-of-classroom experiences like school clubs or homework, we trusted the student's account of the experience rather than joining them for those experiences. Most of the discussion during class and during interviews revolved around in-class activities, which was the main way Mr. Buford had integrated computation into his physics class.

An important part of our methodology is to highlight the perspectives of students, who experience computation-integrated physics firsthand. It is their perspectives on Mr. Buford's curriculum that this paper is about. We intend for our emphasis on participant interpretation to be coupled with a detailed discussion of the research context in which our participants operate. In the next section, we will outline the context of our study and introduce the teacher in whose classroom we generated our data. The rich contextual description we believe is important for practitioners to relate their own experience to and for researchers to understand the setting in which our case study played out.

## III. CONTEXT

Mr. Buford teaches physics at Mulberry High School (pseudonym), a suburban, affluent, racially diverse public high school. He has been teaching at Mulberry for 30 years. In an interview with Mr. Buford, he commented that he tends to try to lean his teaching style towards problem-solving and exploration while still covering the material for the AP physics exams, which he estimates around half of his students elect to take for college credit. He said, "I like to try new stuff," and he confessed that he wishes he had more time to do wide-open, curiosity-driven activities in class: "I think I don't do enough of, 'Okay, so here's this principle that you're responsible for. Today we're going to take some time, and you guys are going to brainstorm an experimental design.'"

One of the recent initiatives that Mr. Buford tried to introduce was computation. He was inspired in part by an existing computation-integrated introductory physics curriculum at Michigan State University (MSU) called Projects and Practices in Physics (P-Cubed) [59]. He began near the end of the 2017–18 academic year by going through the major physics concepts after the AP Exam. For each concept, he recalled, "I think about, does this one seem like it's compatible with writing code to illustrate. Then I try to come up with a scenario, and this is just piggybacking on the scenarios that are used in P-Cubed." For him, the computational activities were meant to be visual, and he used the GLOWSCRIPT programming language [60] along with a minimally working program to do this. A minimally working program [61] is a piece of starter code that will compile without errors and create a visual; however, there are lines of code that need to be edited or added by students to create a realistic physical model. For example, Mr. Buford once introduced a program that showed particles passing through an optical lens without refracting, shown in Fig. 1. The task was for the students to break down their

```
1  GlowScript 2.9 VPython
2
3  scene.fov = .03
4  opax=box(pos=vec(0,0,0), size=vec(1000,3,10), color=color.white, opacity=0.5)
5  lens=box(pos=vec(0,0,0), size=vec(3,200,10), color=color.cyan, opacity=0.5)
6  point1=sphere(pos=vector(-500,80,0), radius=3, color=color.red, make_trail=True)
7  point2=sphere(pos=vector(-500,40,0), radius=3, color=color.red, make_trail=True)
8  point3=sphere(pos=vector(-500,0,0), radius=3, color=color.red, make_trail=True)
9  point4=sphere(pos=vector(-500,-40,0), radius=3, color=color.red, make_trail=True)
10 point5=sphere(pos=vector(-500,-80,0), radius=3, color=color.red, make_trail=True)
11
12 f = 100
13 vel = vector(10,0,0)
14 dt = 0.1
15 fpt1 = sphere(pos=vector(f,0,0), radius=6, color=color.green)
16 fpt2 = sphere(pos=vector(-f,0,0), radius=6, color=color.green)
17
18 while point3.pos.x<500:
19     rate(240)
20     point1.pos = point1.pos + vel*dt
21     point2.pos = point2.pos + vel*dt
22     point3.pos = point3.pos + vel*dt
23     point4.pos = point4.pos + vel*dt
24     point5.pos = point5.pos + vel*dt
25
```

FIG. 1. Starter code and snapshot of the dynamic visualization for Mr. Buford's converging lens activity. Lines 4–10 of the code initialize the lens, x axis, and light particles. Lines 12–16 initialize the focal points, give the particles a velocity, and define a time step for each frame of the animation. Lines 18–24 iteratively move the particles using their velocity and the time step. The snapshot of the visualization shows that the particles pass through the lens in a straight line, indicating that the program does not fully model the physics properly.

understanding of optics into steps so they could edit the computer program accordingly and get the particles to refract, shown in Fig. 2. Mr. Buford would generally begin the computational activities by explaining the minimally working program to the entire class. He would also explain what the output of the code should look like when completed by either running a solution code or drawing the output on the whiteboard. After Mr. Buford finished this explanation, he distributed the program and students were free to work together to create computational solutions.

During the summer of 2018, Mr. Buford attended a workshop at MSU entitled Integrating Computation in Science Across Michigan (ICSAM), funded by an NSF grant with the same name. The week-long workshop was designed to support high school teachers who wish to integrate computation into their physics classrooms.

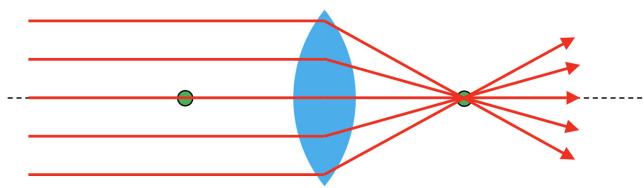

FIG. 2. Diagram of light particles passing through a converging lens. This is similar to the diagram Mr. Buford drew on the whiteboard during his explanation of the activity. Students learned about how light travels through a converging lens during a previous class period.

During the workshop Mr. Buford collaborated with other teachers and facilitators on learning to do physics with GLOWSCRIPT, and by the end of the week, he made a personalized plan for integrating computation into his curriculum for the upcoming year. While Mr. Buford had begun integrating computation at the end of the previous year, he began using it on a regular monthly basis in his AP Physics 1 and AP Physics 2 classes in Fall 2018.

Mr. Buford described in his interview how the computational activities would unfold in class.

Mr. Buford: Grab a laptop and fire it up, and then I go through maybe five minutes—I try to keep it as short as possible—a little explanation of what we're doing, and tell [the students] where to get the starter code and put it in GLOWSCRIPT and start working.

Generally, Mr. Buford would project the minimally working program, or starter code, which he wrote himself, up onto the whiteboard, so students could see as he read through the program's code. Then he explained how important bits of the program worked, ran the program to show the visual at its minimally working stage, and described how it would need to change, occasionally drawing parts of his explanation with diagrams on the whiteboard. Sometimes, he will take a couple minutes near the end of class to project his solution on the whiteboard, so that he can explain a possible solution path. Even though Mr. Buford was showing his own solution on the whiteboard, he would always emphasize that many different solutions exist to the coding projects.

When designing the computational activities, Mr. Buford's approach was to build in checkpoints that students can reach, even if their solutions depart from what he might have in mind. "The ideal to strive for is, 'Okay, now that you've done that, now do this,' and actually have several of those in the bullpen waiting." When he says this, he is talking about progress students can *see* in the GLOWSCRIPT animation window. In the optics activity for example, students can reach these checkpoints first by causing a light particle to move on screen, and then pass through the lens, and then refract, and then add more particles to the animation. Mr. Buford's aim is for students to progress along these steps so no matter how far they go, they still have some sense of success. His main difficulty with this approach has been, "students who struggle can still be working on that initial problem," meaning the first checkpoint that he described earlier. Some students are not even getting past that first step, so they do not get to experience the scaffolded nature of the activity, or even a little bit of tangible progress.

The process by which Mr. Buford designs these activities is to first write the solution himself, and then take out the bits and pieces that he thinks the students should be able to rewrite.

Mr. Buford: I'll try to think of a scenario that's amusing, at least to me, but still is doable. The physics is right in the ballpark of the physics they're supposed to understand. Then the part that I'm not very good at is how much code do I give them, because I give them some starter code…I'll write code that will do what I want it to do, and then I have to try to pick the parts that I would take out and change… and then have them try to figure out how to make it work.

Thus, Mr. Buford tries to address multiple concerns when writing these activities. He tries to balance how much starter code to give students and how much to leave for the students to do, while at the same time making sure that the difficulty and physics content of the problems are appropriate.

Mr. Buford also made some design choices around *when* the computational activities feature in the curriculum.

> Mr. Buford: Those coding activities are culminating activities to studying a concept…It's usually after we've talked about something for a few days or worked on something for a few days. We'll do a coding activity if it fits.
>
> Interviewer: Is that intentional, to have it be after they've learned the concept in part?
>
> Mr. Buford: Yeah…could you use it as a way of developing concepts? I think you probably could. I just haven't done that. I haven't used it that way.

The computational activities in Mr. Buford's class are designed to wrap up a unit. Students have already spent several days learning about a concept, and then Mr. Buford inserts a computational activity. He does not use the computation activities to introduce new ideas, rather they are used to reinforce what students have already learned and to apply those ideas in a new way.

When asked to expand on his views towards computation at the end of the unit, Mr. Buford talked about the importance of visual modeling and coding skills:

> Mr. Buford: I hope it just enhances them thinking about the physics concept that we're trying to learn, ideally… I feel like when you're writing the code for this, you have to understand how projectile motion works, or you can't write code that models that very well… I guess my hope is that that's what we're doing is reinforcing the concepts, and at the same time I just think writing code is just a skill that's so valuable in lots of other areas besides just physics.

Mr. Buford wanted the computation to serve as a way to enhance and reinforce conceptual understanding of physics. His belief is that figuring out the computational activity entails figuring out the physics within it.

On a separate thread, Mr. Buford wanted the computation to serve as a way for his students to learn a skill that is widely applicable outside the realm of physics.

> Mr. Buford: This computational modeling is so appealing to me. It's new. I'm not an expert programmer. I have students that are really good at it. It's cool to see what they come up with and how they come up with it. From my perspective, the problem-solving aspect of that I think is really valuable. The organization and the logic behind it, oh, my gosh. I think those skills are fantastic to have.

From Mr. Buford's perspective, these activities were about more than just physics; they were about building new skills and letting his students' creativity shine. Mr. Buford chose to not grade the activities:

> Mr. Buford: It's okay to not have a grade assigned to every activity in your class, especially with students that are in advanced classes. You don't have to get something for every little bit of effort that you make, so it can be its own reward.

He believed that the opportunity to play with the program and create something intrinsically rewarding was enough motivation for his students.

Overall, Mr. Buford designed the computational activities for the ends of units, when students could reinforce their physics knowledge by applying it to something new and exercise creativity by exploring the computational environment without pressure to turn in a solution. He viewed the computational activities as reinforcements of conceptual knowledge but also opportunities to build crucial computational skills for the future. The computational conditions that Mr. Buford created in his classroom set up the environment that his students were working in and informed the perspectives from students that follow in this study. We include Mr. Buford's perspective here to help readers understand some of the driving forces behind the development of this instance of computational integration. In the sections below, we focus our investigation on the perspectives of Mr. Buford's students, who are the only ones that can tell us how these newly integrated computational activities affect their feelings about themselves and their learning in this context.

## IV. METHODS

We begin our methods section by introducing our student participants, who will be the main focus of our study. The students were selected to represent a broad range of prior experiences (in terms of physics classes and computational exposure) and in-class experience (determined through in-class observations). The aim was not to generalize our results to any sort of population. Rather, we chose a diverse set of research participants because we wanted to describe the variety of challenges students faced in Mr. Buford's class. The class we focused on in this study was Mr. Buford's AP Physics 2 in the 2018–19 academic year. To ensure we respected how the students wished to be represented in this study [62,63], we asked the students after data generation to select a pseudonym and self-describe their gender identity, racial identity, and preferred pronouns.

Otto (he/him) was a junior at Mulberry High School, and he took "regular" Physics 1 with a different teacher before enrolling in AP Physics 2 with Mr. Buford. He always felt behind and that this put him at a disadvantage when it came to the computational activities with GLOWSCRIPT, because he did not have any background with the language. While he did take AP Computer Science the year before, Otto often felt frustrated that his computational background did not seem to help rather than feeling prepared for GLOWSCRIPT. Despite his difficulties with GLOWSCRIPT, he did well in the class, and tended to approach computational activities with the stance that he could just ask Mr. Buford as many questions as it took to figure it out. He usually worked together with Blaine, who also did not take AP Physics 1. Otto self-identified as a white man.

Circe (she/her) was a junior at Mulberry and took AP Physics 1 with Mr. Buford the year before. She usually worked in a large group of six to eight other students who took AP Physics 1 together, including Beck and Ed, and felt a strong sense of community in the class. Often, Circe felt that the computational activities were too hard to authentically engage in, so she usually ended up copying someone else's code toward the end of the period and passing on a working program to someone else, calling it a "copy train." Other than AP Physics 1, Circe had no prior experience with programming, and she did not feel like she was "cut out" for programming or for physics. Despite this, she gave a poster presentation with a couple other students at the state capital about the cool things you can do in physics with GLOWSCRIPT. Circe self-identified as a cisgender Central Asian woman.

Beck (he/him) was a junior at Mulberry, and he took AP Physics 1 with Mr. Buford the year before. He worked in the same large group as Circe, which was usually formed at the start of class with students dragging three tables together. Beck was an avid coder, and he decided to learn more GLOWSCRIPT and do Khan academy physics over the summer after taking AP Physics 1. His dad was a computer scientist. Beck felt that the computational activities helped him understand physics concepts better because it was like "explaining it to the computer." Because he could finish most or all a computational activity without help and he liked to share his code and explain his thinking to other students, Beck was often a resource for other students. Because of his relatively uniform positivity with the computational activities, he did not discuss challenges with much depth. He did, however, describe many positive aspects of computation. As a result, he does not feature in the next section on challenges but does in later sections of the paper. Beck self-identified as a white cisgender man.

Blaine (he/him) was a junior at Mulberry, and he took "regular" Physics 1 together with Otto before enrolling in Mr. Buford's AP Physics 2. He took a helpless stance towards the computational activities, and he was never able to finish an activity during the class period. During one class, he threw his hands up and said, "what's the point of learning code? I can draw this on a piece of paper in fifteen seconds." He often sat with Otto when doing computational activities and he frequently expressed apathy towards programming. His only prior experience working with computer code was when he spent a summer in middle school with his uncle, who worked at a university. Blaine would try to work through programming tutorials while his uncle worked, but he felt like he did not really understand any of it. Blaine self-identified as a cisgender biracial (Black and white) man.

Joyce (she/her) was a junior at Mulberry, and she took AP Physics 1 with Mr. Buford the year before. She usually worked by herself but she also socialized with the larger table, especially after she was done working and ready to share her solution or answer questions. Joyce always finished the computational activity and was often the first in the class to do so. As a result, she spent a lot of time explaining her ideas to other students after she was done. Despite this role, she viewed herself as an average programmer, arguing that she could not solve the problems "in five minutes." She was enrolled in AP Computer Science at the same time and thought that the conceptual ideas from her computer science class helped her when she was using GLOWSCRIPT. Joyce self-identified as a cisgender Asian woman.

Ed (she/they) was a junior at Mulberry, and she took AP Physics 1 with Mr. Buford the year before. She had some additional prior programming experience from participating in robotics club competitions and writing instructions in code for the robots. Typically, she worked in the large group with Circe and Beck, and she tried to figure out and understand the computational activities, opting to ask for help from Mr. Buford or peers rather than join the "copy train" when she got stuck. She said in her interview that she was able to figure out the computational activities around one-third of the time, and this made her feel like she had the ability to successfully program every time. She also felt a strong sense of community in the class. Ed self-identified as a Black agender person. She clarified that she goes by she/they pronouns and suggested for us to pick one to use or alternate between she and they. We opted to use she/her pronouns alone for consistency.

### A. Data generation and transcription

We developed interview protocols and conducted semi-structured interviews [64] with the above six students in Mr. Buford's AP Physics 2 class. The interview questions were aimed to elicit and discuss their feelings about physics class and computational activities in accordance with our research question. The original interview protocol for students is provided in the Appendix. We also interviewed Mr. Buford for the context in the previous section, we took field notes during classroom observations, and we recorded two groups of students working on a computational activity during one class period. The data sources are summarized in Table I. In this study we focused our analysis on excerpts from the six student interviews. It should be noted that we sometimes used in-class occurrences and things that students did or said during the computational activities as prompts during the interviews. Each student's interview was treated as an "anchor point" [54] through which to view challenges from a student's perspective.

TABLE I. Four types of data sources: Student interviews, a teacher interview, field notes, and classroom recordings.

| Data sources | |
| --- | --- |
| Student interviews | Six interviews and three follow-up interviews (follow ups with Otto, Circe, and Joyce) |
| Teacher interview | One interview |
| Field notes | Six class periods |
| Classroom recordings | Two group recordings during one class period, capturing all participants except Circe and Joyce |

The interviews were transcribed for utterances. This choice was driven by a focus on what participants said about their experience, which aligns with our choice to use interpretivist case study. The interviews were conducted to ask about the perspectives of the research participants, and their comments are taken to represent those perspectives. We understand that interview comments can only *represent* how someone feels about their experiences [65], but still we foreground what the participants said, because their responses were prompted verbally. We included nonverbal communication in the interview transcripts when it added meaning on its own to what a student said, such as a face palm or eye roll.

### B. Data analysis

To analyze the interview transcripts, we identified episodes from each interview where the discussion centered around computation, physics, or feelings the student had towards the related classroom activities. It turned out that each interview yielded ten to fifteen episodes of one to two minutes each. The goal with chunking our data like this was to group utterances together into comprehensive statements from the students about their experiences with physics. We carried out analysis on these episodes by taking notes on the episodes one by one, and then tracing out patterns across the different episodes and interviews, treating each interview as a separate data source from which to view a given pattern. We named each pattern according to the common experience or challenge that it represented for students. These names dictated our organization of the first findings section (Sec. V). After outlining and describing the student-perceived challenges, we discuss how the challenges relate to affective constructs, such as mindset, self-efficacy, and self-concept.

## V. STUDENT-PERCEIVED CHALLENGES

We explore the question, *What student-perceived, affect-based challenges do high schoolers face in computation-integrated physics?* by presenting the interview data in which our high school student participants described their experiences and feelings around doing computation in their physics class. In the results below, we describe patterns in the data that constitute different affective challenges that students faced when doing computation in Mr. Buford's class. The challenges listed below are in no way exhaustive, nor are they necessarily confined to computation-based settings, but instead represent an initial set of challenges experienced by students in this context. In the order presented, we address each challenge: Stress or frustration, feeling worse at physics, unbelonging and stereotypes, repeated confusion, interpreting code, and interpretations of implementation.

### A. Stress or frustration

One of the main challenges posed was the additional stress that computational activities brought to students in Mr. Buford's class. Stress often accompanies new experiences but what made this a challenge was that students often saw the stress as uncalled for. They felt that they already knew the relevant physics concepts, and computation was just forcing them to jump through hoops in order to translate their physics knowledge into code. These experiences were often accompanied by frustration when difficulty was unexpected. The unexpected frustration and the unnecessary stress combined to make some students feel unprepared and inclined to give up.

When Circe talked about stress in the interview, she spoke more generally about the stress she felt during all computational activities and coping strategies she employed.

> Circe: I feel like it's just unnecessary stress, and I'm not about to put myself through that. So I just kind of sit there with the people, and we just talk and wait for one person to figure it out. Like I said, a copy train.

She felt stressed out during the computation, and her reaction was to not "put herself through that." Rather than confront the difficulty and "unnecessary stress" head-on, she opted to copy answers along with the rest of the group. Her response was to disengage, indicating either that she did not believe she could figure it out or that the stress of sticking it out was not worth it.

At another point in her interview, Circe talked about how the computational activities, or "code" as she put it, frustrated her. During this discussion the interviewer asked a question to get an explanation for what she meant.

> Interviewer: What about the code frustrates you?
> Circe: It's like, you think that you should do a certain thing, input a certain value, or a new part of the thing, and you do that, and it's just completely wrong.

> And you sit there and you're like, okay, well, freak you, coding!

Circe felt that even when she made everything right in the computer program, or seemingly right, it ended up being completely wrong. In this way, there was no middle ground when it came to computation, and this made her feel that she could not do anything right during the activities. Her reaction was anger ("freak you!") towards computation. There was no resolution, only frustration and giving up.

Another student, Ed, also discussed experiencing significant stress, but she did not disengage as readily as Circe did. Ed's stress was also "undue" as she said below, and it had to do with a tension between the computation and Ed's perceived physics knowledge.

> Ed: I feel like [computation] causes me, sometimes, a lot of undue stress, which is like 'Oh, you don't know this and this and this.' So it's like, 'you do, you just think about it in a different way, but that's not a way that can be programmed on the platform.'

She felt stressed because of how the computation challenged what she thought of her physics knowledge. The stress was associated with the feeling of not knowing, and she had to coach herself out of the difficult feeling essentially by saying, "you *do* know physics, it's the computation that's confusing." The "undue"ness of the stress made it seem as if Ed viewed computation-integrated physics as unexpected in relation to the physics she was used to, and the stress itself was tied to these expectations, because she *did* feel like she got it when it was just physics without computation.

Ed also felt some unpreparedness for the computational activities. When asked about whether she saw herself as "good at the coding activities," she responded by commenting on the frustrations of seeing the physics content being stripped of its familiarity.

> Interviewer: Do you think you're good at the coding activities?
>
> Ed: Not really, actually, which is kind of sad for me to be honest, because you have this interest in something, but it's back to why physics is so frustrating, because it's something that's like "Oh, this is familiar, I know this," but then it's just slightly slanted a little and just becomes, because you expect it to be this way so much, when it's this way, it's just, you can't handle it.

She linked her negative self-evaluation to a frustration about physics in general. She compared her computational frustration to the common experience of learning physics concepts that seem to defy intuition about how the everyday physical world works. Computation made familiar material confusing for her. Though the stress functioned in a different way than it did for Circe, the common thread was that it came from the computation. Ed felt like she built expectations for how her ideas would play out in GLOWSCRIPT, but it never seemed to work out—she could not "handle it." From this example, we see that Ed dealt with her frustration by separating her conception of physics, which was familiar and understandable, from the computational activities, which defied her expectations and caused her stress.

For Circe and Ed, computation added an extra, needless stress. Their reaction was to find ways to avoid the stress. For Circe, this meant copying others' solutions. For Ed, this meant separating physics and computation mentally as a defense to preserve her self-view as a competent physics student. Other students also experienced stress but did not articulate it in these terms, such as Blaine becoming apathetic towards computation after repeatedly getting stuck or Otto feeling stumped and behind because of his lack of previous GLOWSCRIPT experience. Both of these accounts are described further in the challenges below.

### B. Feeling worse at physics

Another challenge students faced was the way that computation seemed to test and even diminish the strength of their perceived physics knowledge. This is not necessarily a bad feature. After all, Mr. Buford wanted the computation to "enhance them thinking about the physics concept that we're trying to learn… I guess my hope is that that's what we're doing is reinforcing the concepts." For some students, the "enhancement" of physics thinking instead meant that they had to reconsider what they knew for the purposes of the computational activity, and this reconsideration often led to feelings of incompetence at either physics or computation. An example of this challenge is when Ed felt "undue stress" in the previous subsection. She recalled thinking, " 'Oh, you don't know this and this and this…You do, you just think about it in a different way, but that's not a way that can be programmed on the platform.' " She told herself that she *did* know the relevant physics, just not in a computational way. In effect, she separated the two domains (computation and physics) in her mind, so that her difficulty with computation would not affect her view of her physics competence.

Later in her interview, Ed reflected on how she viewed the connection between computation and physics. She even suggested that computation changed her view of her physics knowledge.

> Ed: I think coding definitely affects my perception of my own knowledge about physics… GLOWSCRIPT especially, I feel like it caters to a very specific kind of learner, a very specific way of learning physics…it just requires you to take apart the numbers in a very strange way. Well, it's not a strange way, it's a strange way *for me*.

She felt that being good at computation (especially GLOWSCRIPT-based computation) was like being good at learning physics in a special way. Ed felt unable to learn in this "strange way." When she struggled with computation, it felt like the class had been redesigned with a different type of physics learning, and Ed's physics knowledge did not line up with the "very specific way of learning physics."

For Joyce, *getting stuck* during computational activities is what made her question her physics ability. Her self-doubts about her physics knowledge were rooted in not being able to translate the formulas she knew into code.

Joyce: Sometimes it's made me think that I'm not as good at physics because when you do everything that seems right on there, or if you use that equation, you get the right answer on your own, but you can't program it, then that made me feel challenging.

Joyce linked her GLOWSCRIPT-based struggles to feeling bad at physics. This happened when she felt like she programmed everything right and she knew how to do the problem on paper, but it still did not work on the computer.

The challenges that Joyce and Ed reference in the interview excerpts are not necessarily a bad thing—in fact it might be a sign of growth and learning that they are being forced to reconsider their physics knowledge in a way that aligns better with computational demands (assuming these computational demands are part of an equitable learning environment). However, these experiences are challenges all the same and must be addressed because they pose real concerns for students. For both Ed and Joyce, computation forced them to reconsider their physics competency because they felt incompetent when doing physics with computation. We do not have the data to say whether or not these feelings of incompetence were temporary, but it is clear that they constituted real affect-based challenges when doing computational activities. Some students who experienced such feelings—or even stress and frustration like in Sec. VA—struggled with a tension between their self-views of their computational competence and physics competence. For some students who found computation to be unexpectedly hard, an appealing narrative could be, "I'm good at physics already, this is just me being bad at computation." It is much harder to swallow the pill labeled, "I'm not as good at physics as I thought."

### C. Unbelonging and stereotypes

The feeling of not belonging in computation and/or physics was also present in Mr. Buford's classroom. This challenge is not necessarily brought on by the implementation of a new curriculum, but difficulty with the learning materials can exacerbate existing feelings of exclusion. Furthermore, computation-integrated physics is the intersection of two STEM fields (computing and physics) that have struggled to achieve diverse participation from people with different identities, such as women, people of color, people with disabilities, LGBTQ + people, and people of lower socioeconomic class [7,66]. As an example of a student feeling out place, we look to Circe, who talked at length about this when she thought about the computation in Mr. Buford's class. In the excerpt below, Circe noticed patterns among her peers related to computation and physics. She used the word "coding" to refer to the computational activities.

Circe: I think I've noticed that there's people who are really good at physics that are also really good at coding. I think there's a pattern there. I have a lot of friends who are really good at coding, and they're usually really good at physics, and vice versa. It's like, I don't know. I guess it's all the same kind of brain.

Circe compared being good at physics to being good at computation. She had noticed that a lot of her friends were good at both, and there seemed to be a connection. It is "the same kind of brain," she said, which indicates that she viewed those peers' academic abilities as intrinsic qualities that they had. The language she used suggested that she saw herself on the outside of this peer group: "I've noticed that there's people," "friends," "they." By using otherizing language, she positioned herself as *not* having the same type of brain, indicating that she saw herself as not naturally cut out for physics and computation like a lot of her peers seemed to be.

Later in her interview, the conversation again turned to her sense of belonging in physics. Circe had established earlier that she was not interested in pursuing physics after high school, but she went on to imply that computation somewhat confirmed her thinking.

Circe: I don't know if coding makes me feel like I don't belong in physics. It doesn't make me feel like I do belong in physics.

She was sure that computation was not making her want to be a part of the physics community. Even though computation might have been integrated into the course as a way of making physics more authentic to students, its effect on Circe was not beneficial to her sense of belonging.

In naming and characterizing the challenge of not feeling cut out, we acknowledge that many students choose to leave physics, and this choice can be in line with their interests and based on a realistic understanding of what it means to do physics and be a part of the physics community. However, many students can build views of physics or computation based on stereotypes of who does physics and unrealistic views of what physicists do [67]. One possibility, based on Circe's views about the "kind of brain" that is made for physics, is that she bought into some of these stereotypes, particularly to be good at physics you must be an innate "physics genius" [68].

Similarly, we see stereotypes of programmers and programming show up in the classroom. We say "programming" here because often the students who adopt these stereotypes do not distinguish between the computational activities (where student program physics) and more general programming. An episode that encapsulates this view is when Joyce discussed why she felt like an average student despite her repeated success at the computational activities.

Joyce: I think I'm better than average, which is someone who doesn't know how to code at all. But I'm not…

I can't just look at the scenario and just code it in five minutes. I'm definitely not that kind of person. I don't know. Just average I guess.

Joyce believed she was average compared to all programmers, implying that people who can look at the problem and do it in "five minutes" are the good programmers. None of her physics classmates were this fast, but she compared herself against this imagined programming genius anyways. This led Joyce to feel average despite being one of the most competent programmers in her class.

Stereotypes like the genius, five-minute coder can make computation feel inaccessible, and it can make it hard for students to build a sense of belonging in computation and/or programming. The challenge of stereotypes lies in this perception of unbelonging. The fact that some students must overcome this perception *and* still perform well in class in order to see themselves as computationally competent is a significant barrier.

The integration of computation into physics leaves the physics classroom open to stereotypes about programming *and* computer science. Students have understandings of what it means to contribute to computer code, and sometimes those understandings are built on unrealistic stereotypes about who does programming, what programming looks like, and how people become programmers. This is on top of the stereotypes of what it means to do physics, who gets to do physics [69], and how one can succeed at physics (e.g., "physics genius"). The prospect of computation introducing even more stereotypes into the physics classroom poses a significant challenge.

### D. Repeated confusion

Because of the open-ended nature of the computational problems in Mr. Buford's class, many students had difficulty working on them. For example, there were many places where students were confused, encountered errors, or did not know how to proceed. How students reacted in these moments could lead them to interpret their experiences as failures or could lead them to success with the problem. The ways that students interpreted the successes and failures outlined below constituted an affect-based challenge for some students.

From Otto's experience, he often found success with the computational activities by working through his difficulties and trying to simplify the problem. Even though the activity was confusing to him, he felt like he could make sense out of it after thinking about it. He walked us through his general approach to computation in Mr. Buford's class.

> Otto: When I'm working through it, I'll be like, "this is confusing." And I'll start working through it. I'll try to simplify it to something that I can understand. Then I'll usually be able to think about it and be like, "Yeah, that makes sense. I can implement that."

Otto's strategy to deal with confusion was to simplify the problem until he understood what he needed to do. When he said he was "usually" able to figure it out, he indicated that there was a pattern in his approach to computational activities. The phrase he told himself was, "I can implement that." Whether or not he succeeded, Otto usually came to a point during computational activities when he at least *felt* like he could, even if he started the problem feeling confused. As a specific example, he remembered getting stuck and eventually figuring out a complex computational activity about the motion of charged particles in a magnetic field.

> Otto: There's a part where you had to use vector cross products to show the direction in which it would be moving, from like the direction of…the field and its movement already. That clicked a little bit after I realized how that function worked.

Though he encountered a confusing function, he figured it out. The function in question was the cross product function. His success in getting the function to work and understanding it is evidence of Otto's persistence in face of his typical computation-based confusion.

For Ed, experiences of success were more rare but not unheard of. When she did finish a computational problem it made her feel like she could do *any* of them.

> Ed: On like one out of the three times we coded, each of those one times where I've actually finished the whole thing, that always makes me feel like, "well you finished that one, you can probably do all of these."

Approximately one out of three times, Ed could figure out the code, and it was a big confidence boost. For her, it was the act of completing the program that made her feel the sense of attainment. Though she usually did not finish, on the times that she did, it was a reaffirmation that she had the ability to succeed at doing the computational activities. The intermittent successes sustained her.

Blaine, on the other hand, discussed how he had recently given up on engaging with computational activities because of his failures to achieve anything that he perceived as progress.

> Blaine: I mean, I would try if I could literally get like anything. But since I literally can't get anything but a blank screen, I don't really try to do any more cause I'll put in a hundred things and then I'll just get a blank screen or I'll get some error.

No matter what he tried, Blaine always got the same result: "a blank screen or some error." Both results are associated with a nonworking animation, a fate to which Blaine had resigned himself. Not only was this a wholly negative self-evaluation, but it was also a source of apathy and disengagement for Blaine. He experienced repeated roadblocks, and he came to associate his relationship to computation with incompetence.

> Blaine: I just, I just don't even care. I'm like "whatever dude. I can't do this sh*t."

He felt like he could not do the activities to the point that he just "[didn't] even care" anymore. He provided a sharply negative statement, saying, "I can't do this sh*t." He had no successes with computation, and by the point of the interview he had given up entirely. Blaine was one of the two students who did not take AP Physics 1, so his first exposure to computation-integrated physics was Mr. Buford's class. This points to the importance of having positive experiences and moments of success when learning a new curriculum as suggested by Kinnunen and Simon [32]. Blaine had no memories of success and articulated no hope that he would improve.

Some of these students experienced setbacks or confusion in the computational activities. While Otto persisted through such a moment and eventually figured it out, Blaine interpreted his lack of computational success with feelings of apathy and inability. Ed had enough positive experiences to feel competent, but all the same it was concerning that some of Mr. Buford's students were not having any positive experiences with computation. The prospect of students developing negative views about computation after repeatedly failing at computational tasks presents a unique challenge, especially when these failures are tied up with their first impression of computation-integrated physics.

### E. Interpreting code

Another common challenge was brought on by the need to interpret code and errors in GLOWSCRIPT. This has been previously documented with students learning physics through VPYTHON [70]. Students in Mr. Buford's class often felt that they had a decent understanding of how to use the relevant physics and apply it to the context in which Mr. Buford set up the computational activity. The challenge came when they received an error message or had to interpret or write code to execute their ideas. The elusive meaning of the error message or the challenge of using GLOWSCRIPT syntax was enough to derail the activity for these students.

For Blaine, the computational activity that he described involved modeling rays of light passing through an optical lens. He had trouble with the very first step because he could not figure out how to use GLOWSCRIPT to animate a line to represent the light ray.

Blaine: I feel like I'd like [the computational activities] if I knew what I was doing. I literally wrote (laughing). I literally wrote "line," just like "line period," to try and get a straight line. I don't know anything!

When he talked about what it was like to troubleshoot after getting stuck, he laughed about how little he understood GLOWSCRIPT. He *guessed* at what the proper syntax would be because he did not know any GLOWSCRIPT commands for creating something that looked like a line. He attributed the whole experience to his lack of knowledge: "I don't know anything!" This admission was reaffirmed below when Blaine described his inability to interpret an error message because it referred to "line 17," or the seventeenth line of the computer program, which he was unable to interpret.

Blaine: I'll get some error. "Line 17." Well I don't know! I don't know what line 17 is, man.

In this case, Blaine could not interpret the error message that the computer provided. His responses about "not knowing what line 17 is" and "not knowing anything" indicate that Blaine felt that he just did not know enough about the GLOWSCRIPT language to do computation.

Otto had a similar, though less severe, reaction to getting stuck on using GLOWSCRIPT. He discussed the process of figuring out the relevant physics but not being able to translate his ideas into code.

Otto: The electron moving through the magnetic field… I know what direction it should be moving and everything, how its velocity should be affecting everything. But I don't know how to put that into computer words… Even when I know what should be happening, it just wasn't happening, because I don't know how to use GLOWSCRIPT that well.

He explained the roadblock: "I don't know how to use GLOWSCRIPT that well." Though his programming inexperience prevented him from succeeding, Otto acknowledged that he *did* know the ins and outs of the noncomputational part of the physics problem. He contrasted what he did and did not know, saying, "but I don't know how to put that into computer words." Otto's experience was different from Blaine's because Otto was able to identify what he knew about the problem and what exactly he got stuck on. This shows that the challenge of interpreting code can present differently for different student and different contexts, but in each case it can present a barrier all the same.

Circe also described her challenges with understanding the code. She recalled starting a computational activity and immediately feeling lost.

Circe: I feel like something like coding can't help you understand physics better if you don't understand what the code means in general. He gives us the code to start off with, but none of us really understand what that means. So we look at [the starter code] and we're like, "what does any of that mean?" So then you add things to that, but you don't understand why.

She often felt that she did not understand the program, or starter code, which Mr. Buford distributed to be worked on. This had the perceived effect of preventing Circe from learning physics through computation. She even described attempting to engage with the activity and add her own code but feeling confused and directionless. Her understanding of the computation was that success depended on computational literacy of GLOWSCRIPT and that some students did not have the tools to engage on that level. Her use of "we" indicates that this experience of confusion was shared among her peers and her.

Even for students who had seen programming before, using the GLOWSCRIPT language, structures, and syntax was still a challenge. For example, Otto had taken a physics class and a computer science class before enrolling in Mr. Buford's physics class. Despite these experiences with the "ingredients" of computation-integrated physics, Otto still felt like Mr. Buford's version of computation was new.

> Otto: It's a lot more *physical* in GLOWSCRIPT because in the other class I took with coding, it was more just data and lists and whatever. But this you're having a particle moving through whatever so you have to use like vectors and all that. That's new to me. I haven't done anything involving movement and displays and that.

He said GLOWSCRIPT physics was unique because of the movement and the visual nature of the activity, whereas computer science was about "data and lists." Computation in physics felt totally new to him, from the language (GLOWSCRIPT) to the conceptual features (e.g., vectors, movement, animation). Doing computation with GLOWSCRIPT was different from both physics *and* computer science in Otto's view, and this unfamiliarity made it difficult for him. The difficulty manifested when he had to combine physics with computation: "I know what direction it should be moving and everything, how its velocity should be affecting everything. But I don't know how to put that into computer words."

For Otto, who had prior experience with both physics and computer science, working with GLOWSCRIPT still felt totally new, and he found it difficult to put what he knew into "computer words." This indicates that interpreting code might be a significant challenge for all students to some degree and that prior experiences with code do not directly translate to success at computation-integrated physics activities. Otto pointed to the specific features of the integrated format (making particles move, using vectors, making displayed simulations) that were still a challenge for him. Just because he had the separate physics and computation pieces, it did not mean that Otto felt able to combine them, and he still struggled with translating the ideas into the computer words.

Blaine, Otto, and Circe shared above how they got stuck because of a difficulty with the computer program, not the physics concepts. The impact was twofold. First, it stopped these students in their tracks when they did not know how to deal with code during a computational activity. Second, it caused negative affective responses, like Blaine's self-evaluation ("I don't know anything!") and Circe's indictment of the activity itself ("coding can't help you understand physics better if you don't understand what the code means").

### F. Interpretations of implementation

There were also some implementation-based challenges that students faced in Mr. Buford's class. These were related directly to the students' interpretations of the computational activities and pedagogical choices made by Mr. Buford. We share these not as a critique of Mr. Buford's implementation but as a way to illustrate the variety of challenges that can arise for students and how those can depend on the context.

#### 1. Assessment and motivation

In Mr. Buford's class, the computational activities were intentionally not graded. Mr. Buford felt that because the activities were new and not explicitly a part of the AP curriculum, they could go ungraded and simply serve as opportunities for students to engage with physics concepts more deeply than they normally would. He explicitly said in his interview, "You don't have to get something for every little bit of effort that you make, so it can be its own reward," indicating that he viewed the computational activities as intrinsically motivating.

In the interviews with students, we saw that students understood this motivation and experienced it for themselves at times. For example, Ed expressed a similar view of computation, that the purpose was to get a better grasp on programming concepts, which in turn helped her see the connection between formulas and actual physics phenomena. We provide the excerpt below.

> Ed: And just seeing how just changing a couple of numbers could change the entirety of the coding was interesting… That was helpful for me to get the whole concept of coding.

However, at a different point in the interview, she articulated a much bleaker view of what computation was all about, referencing the grading policy.

> Ed: [Coding activities] are just really tedious. When I'm doing it, I just feel like there's something else I could be doing… I feel like coding is like something you kind of know… and it just feels kind of like busy work, but not busy work that he's going to grade, so it just feels useless.

The goal of computation, as Ed articulated here, was nothing. In her view, because it was not graded, there was no point in engaging. The computation was "tedious…busy work" which made Ed want to disengage even more. Had the activities been graded, she might still have found them tedious, but the fact that they were ungraded meant they were "useless," at least in how Ed viewed them in this moment.

Ed's frustration at computation did not last throughout her interview, but the above excerpt demonstrates that the ungraded nature of computation in Mr. Buford's class can contribute to a feeling that computational activities serve no purpose. Feelings like this can impact students' motivation ("feels useless"), and given the open-ended, ungraded design of many computational problems, motivation was important for students to want to explore the activities.

As Mr. Buford indicated, it was reasonable to not have every single activity be graded or externally motivated.

In fact, we can imagine several arguments for leaving computational activities ungraded. For example, teachers might want to reduce the pressure and stress of grades while students are doing a novel, unfamiliar task. However, as Ed's response indicates, there is a need for messaging about why students are asked to complete an ungraded activity, why the activity is not graded, and why engaging in the activity can still provide benefits to students.

#### 2. Solutions and "right" answers

When introducing the computational activities, Mr. Buford would explain the minimally working program and show students what the output of the code should be when fully working (by either drawing it on the whiteboard or showing the output from his solution code). He intended this as a way to show students what the end product should be in an otherwise open-ended activity. Mr. Buford was careful in his explanations to emphasize that there could be multiple right answers or solution paths to the computational activities.

Despite his caution and explanation of multiple paths, knowing that Mr. Buford had a "correct solution" posed an affective challenge for some of his students. For example, Circe was a student who viewed "success" at the computational activity as "being right," and she said that her own ideas were always "wrong" when it came to computation. Below, the interviewer asked her about this view.

> Interviewer: How do you know it's just wrong?
> Circe: Because you see the answers. I guess there's multiple answers, so you might not be completely wrong…but the one that we're given, or the one that the smartest kid in class figures out is different than the ones that we had.

She articulated that the goal was to get the answer that the teacher had or the smartest kid in class had. Anything else she saw as wrong. She even acknowledged that there could have been multiple solution paths, but she still interpreted a mismatch in her answers as "not completely wrong" and set up this comparison for her work versus a "smartest" or "given" (teacher's) solution. Circe reasoned that "because you see the answers," hers (which did not match) must be wrong.

From this perspective, showing the final output to the class might inhibit students' ability to see paths beyond the one they are shown and might pose an affective challenge for students who need to reckon with the tension between being right and engaging openly with the problem. This desire to be right also can prevent students from exploring the problem setting and making mistakes from which they can learn important aspects of the problem.

That said, we do not know what would have happened if Mr. Buford did not provide the output for the computation problems. Without knowing the output, students could potentially struggle more with interpreting the code or might encounter more confusing moments as they work through the open-ended problems. These implementation-based challenges are directly related to choices that Mr. Buford made in integrating computation into his physics course; however, they do not represent all the challenges related to implementation that students could face. More studies should be done in a variety of contexts that look at students' other implementation-based challenges.

## VI. CONNECTION BETWEEN CHALLENGES AND THEORY

From students' interviews, we showed that they faced a variety of challenges when computation was integrated into their physics class. While it was not the explicit focus of this study, the students' statements point to theoretical constructs in education research that might help better understand students' experiences and how to help address these challenges in the classroom. Specifically, we found ties between students' comments, their mindset, self-concept, and self-efficacy.

Briefly, self-efficacy is a person's belief in their own ability to complete a task [71,72]. Within the context of a computation-integrated physics classroom, self-efficacy would address the question of "how well can I do computation in this physics class?" Mindset, at its simplest, is a person's belief in their ability to change their own traits or competencies [49]; thus, mindset would address the question of "how much can I improve at doing computation?" In contrast, self-concept is "a person's perception of self…inferred from their responses to situations" [73] (p. 411). Rather than being task related (as self-efficacy), self-concept is in relation to an entire subject area. This would address the question of "how is doing computation related to me?" In the sections below, we define in more detail each of these constructs and how they are related to our data. We then discuss the overlaps in these constructs and the implications for instructors and researchers.

### A. Self-efficacy

Originally developed by Bandura, self-efficacy is "concerned with judgments of how well one can execute courses of action required to deal with prospective situations" [72]. In discussing how self-efficacy relates to students, Bandura suggested that it contributes to motivation and confidence within a given academic subject: "The higher the students' beliefs in their efficacy to regulate their motivation and learning activities, the more assured they are in their efficacy to master academic subjects" [71] (p. 18).

Since its introduction, self-efficacy has been broken down into four sources: mastery experiences, vicarious learning, social persuasion, and physiological state [71]. We looked at how the four sources have been used in STEM education research to gain a deeper view of what they could mean for a computation-integrated physics context [74,75]. Mastery experiences refer to the impact of successes and failure: "successes heighten perceived self-efficacy; repeated failures lower it, especially if failures occur early in the course of events and do not reflect lack of effort or adverse external circumstances" [71]. In our case, completing a coding task could count as a mastery experience, or receiving an error message from the coding program could be seen as a "failure." Vicarious learning is when a student makes an adjustment to their self-efficacy after witnessing a peer's performance. For example, a peer's success at a computational task can raise self-efficacy if the student then thinks they can succeed too, but seeing a peer fail despite effort can lower the observer's self-efficacy for related computational tasks. Social persuasion is about external appraisals of ability that a student then internalizes into their self-efficacy. Evaluations can come from peers, authority figures, or other participants in the domain where the student must perform. Social persuasion need not be verbal or direct, and its effect depends mainly on how the student perceives it. Physiological state refers mainly to stress "as an ominous sign of vulnerability to dysfunction" [71]. Students, when they are stressed, expect to perform worse, whereas when they are calm and clear-headed they might feel a boost to self-efficacy.

A few examples from computation education research show how self-efficacy can be used in computational settings and how it can reveal information about student learning. Self-efficacy was employed by Lishinski et al. [30], who viewed self-efficacy as a reciprocal feedback loop, where self-efficacy judgments based on affective responses can have a long term effect on learning outcomes. The authors found that previous programming experiences impacted future performance in part due the effect that past experiences had on self-efficacy, whether positive or negative. Kinnunen and Simon [31] used self-efficacy to describe students' affective responses to a computational assignment in an introductory-level university computer science class. When students made an affective self-assessment, the authors were able to describe it in terms of self-efficacy, indicating a connection between self-efficacy and the act of affect-based evaluations of oneself. In a follow-up study [32], Kinnunen and Simon used the four sources [71] to understand how self-efficacy was tied to experiences that students had in the course. They also considered in their framework how self-efficacy could evolve in response to experiences and what could set this evolution in motion. A year later, the same authors [33] returned to self-efficacy, this time using it to describe emotionally charged events they observed where students evaluated their own abilities and consequently altered or reinforced their self-efficacy for programming. The evolution of how Kinnunen and Simon [31–33] used self-efficacy to explore programming experiences demonstrates a precedent for connecting self-efficacy (and its sources) to computation.

We can see these sources of self-efficacy in our data, with examples that might be either contributing to or degrading students' self-efficacy in computation. For example, in Sec. V D, we saw Blaine, Ed, and Otto take on very different responses when faced with confusion and uncertainty in the coding activities. Otto demonstrated a persistence in his approach to the problems, experienced multiple successes (mastery experiences) with the computation problems, and often said high self-efficacy statements like "I can implement that." In contrast, Blaine experienced very few mastery experiences. This connects to several of his statements which aligned with a lack of computational self-efficacy. He said, "I can't do this sh*t" and "I don't really try to do any more 'cause I'll put in a hundred things and then I'll just get a blank screen or I'll get some error," which directly tie his lack of success ("blank screen" or "get some error") to his belief that he cannot code or cannot make progress. That said, Ed's experience demonstrated that mastery experiences do not have to be all or nothing. Ed had some moments of success with the code, but she indicated that it was only one in three activities. However, even those moments of success made her feel like she could code and contributed to her belief that "you finished that one, you can probably do all of these." All three of these students pointed to the importance of mastery experiences in building views related to self-efficacy, especially Ed's case, which highlighted that not all computational experiences need to be successful.

There were also indications of the other sources of self-efficacy in our data. For example, Joyce referenced the stereotype of a "fast coder" in her statements in Sec. V C, saying that she was simply average because she could not "just look at the scenario and just code it in five minutes." Even though Mr. Buford never set any expectations about how fast students were expected to code, Joyce still had this idea that the good coders were able to just look at the code and do it. Research has shown that perceptions like these can come from societal stereotypes, media portrayals of programmers, interactions with peers, and other forms of social persuasion [69,76–78]. Social comparison of programming speed has been shown to reduce self-efficacy [76]. Ultimately, this perception encompassed how Joyce saw herself and how she evaluated her skill. We also showed that Circe and Ed described computation as a stressful, frustrating activity in Sec. V A. This outlines one of the physiological states that can contribute to self-efficacy. If a student's experiences of coding are all taking place in a highly stressful, tense physiological state, then that reduces their self-efficacy and garners a feeling of inability to complete the task. We saw this with Circe, who directly stated that she's "not going to put [herself] through that" because the programming is "just unnecessary stress."

The sources of self-efficacy open questions for additional research in computation-integrated classrooms. For example, what tasks and what grain-size lead to mastery experiences? Does interpreting an error message successfully count as a mastery experience or does the whole program have to be completed for students to feel successful? How can we as instructors and facilitators help students see their success in each of these moments? How can we help students approach computation without a stressful physiological response, while at the same time not seeing computation as "useless" or "busy work?" At this point, we do not have answers to these questions, but our results from the challenges students face would indicate that more research is needed in this area.

### B. Mindset

Dweck [49] defined mindset in terms of self-beliefs about the mutability of abilities and delineated between fixed mindsets and growth mindsets. She argued that a fixed mindset is detrimental to learning because students who embrace this mindset lose motivation more easily and they are harsher judges of self when faced with adversity. On the other hand, students who embrace a growth mindset build motivation to improve when they experience failures. Blackwell et al. [79] provided a review of perspectives a student would hold depending on how their statements and actions aligned with mindset. The most fundamental perspective is that growth mindset aligns with a belief that one can improve their intelligence through effort, whereas fixed mindset relates to believing that intelligence is unchangeable. Growth mindset is about studying to learn, seeing mistakes as learning opportunities, believing that effort is good because it makes you smarter, and seeing knowledge as something that can be worked for [49,79]. Fixed mindset is about studying to prove smarts or superiority, avoiding mistakes for fear of being seen as stupid, believing that too much effort signifies lack of intelligence, and seeing knowledge as something that comes from authority figures [49,79]. When students fail, some may react in ways aligned with growth mindset, believing they need to change their studying strategies. Some students may react to failure in ways aligned with fixed mindset, believing they failed because they are stupid or because the assessment was unfair. As a disclaimer, the theory of mindset is flexible, meaning that reacting in a "fixed mindset" way does not mean one will always react in that fashion [49]. Also, mindsets can vary between contexts or even within a single context, meaning people can hold views related to both growth and fixed mindset about different subject matters or even at the same time [49].

From the literature, mindset has been used in some initial studies to describe students' approaches to computation. In one study, Scott and Ghinea [80] set out to discover whether programming-specific mindset could be differentiated from general mindset for school. They discovered that the unique nature of programming activities led students to embrace a specific mindset for programming, different from a more general, school-based mindset. To track learning in connection with mindset, an intervention study was devised by Cutts et al. [81]. They intervened in an introductory university programming class by having tutors teach mindset-related strategies. The issue of getting stuck was focal: the students' mindset-related views hinged on whether they attributed being stuck to internal factors (leading to an embrace of fixed mindset) or external factors (leading to an embrace of growth mindset). These findings suggest that mindset-related views could change or even develop anew when computation gets introduced into a physics curriculum. Lodi [82] performed a similar study to Cutts et al. [81], but he focused on high school students and sought to understand how the computer science curriculum impacted mindset-related views. He argued that students with learning-oriented goals (e.g., aiming to learn and be challenged) aligned their views with growth mindset, whereas students with performance oriented goals (e.g., aiming to score well and avoid challenges) aligned their views with fixed mindset. These studies highlighted some of the same features of mindset that emerged from Dweck [49] and Blackwell et al. [79], which gives us precedent for applying these theories to a computational education setting.

In our data, we saw similar perspectives mirrored in how students articulated challenges in Mr. Buford's class. For example, in Sec. V F, Circe recognized certain answers as "right," and those answers came from the teacher or the smartest students in class. This aligns with the fixed mindset tendency to look to authority or expert figures (like teachers) as the only trusted source of knowledge. Tendencies of people to value accomplishments and grades because they signify high intelligence align with aspects of fixed mindset, whereas tendencies to value learning because of its connection to *improving* intelligence align with aspects of growth mindset. Circe articulated a tendency to consult the teacher's solution to see if hers was right, which represents a potential challenge in other settings where computational activities are designed to have multiple solutions and unanswered questions built into the learning process. For students who embrace a fixed mindset at times, this design could present significant barriers to success.

Another example comes from Secs. V C and V B, where we observed Circe and Ed provide similar views about feeling out of place or not knowing how to proceed when confronted with computational challenges. For Circe, feeling out of place was tied with her belief that being "really good at physics and coding" meant having "the same kind of brain." When students take up the view that they need to be built a certain way in order to succeed at physics and/or computation, they align their views with fixed mindset, which at its core says that intelligence is an inherent characteristic and impossible to change. For Ed, she felt that her understanding of physics was questioned or alienated when she had to do physics with computational tools, to the point that she believed she "just [thought] about [the material] in a different way," and she emphasized the computation was only strange *for her*. This distancing that Ed does indicates that the challenge was related to fixed mindset, because she attributed her difficulties to her self-perceived faulty way of thinking, and she viewed computational learning as "catered to a specific kind of learner," distancing herself from the opportunity for *her* to learn during those activities.

Lastly, we return to Sec. V D to compare the mindsets that described what Otto and Blaine said when faced with confusion. We focus on their difference in persistence. Both students articulated getting confused or stuck, but Otto's response was to embrace the challenge ("I'll start working through it, I'll try to simplify it"), whereas Blaine's response was to give up ("I don't really try to do any more"). For Otto, the setback was an opportunity to learn, which aligns with growth mindset, whereas for Blaine, the setback was paralyzing, which aligns with fixed mindset. The contrast between how students respond to these challenges is closely aligned with mindset theory, which indicates that mindset can be key in explaining whether students succeed at overcoming challenges in Mr. Buford's computational activities.

Our work suggests building on the premise that mindset is linked to how students respond to computational challenges. A follow-up study to this one found that mindset described students' statements and actions in far more detail than we can articulate here [83]. Other questions could take this work further, for example, how do students develop views related to mindset theory in their computational work? Are there pivotal experiences (like mastery experiences for self-efficacy) that impact students' mindsets in significant ways? Our data would also suggest observing how students treat computational challenges differently in the wake of mindset interventions, similar to many others' recommendations [84–88]. We also recommend studies on designing opportunities for students to embrace growth mindset could help students in other ways in a computation-integrated physics context. We do not have the answers, but our results from the challenges students face would indicate that more research is needed in this area.

### C. Self-concept

Shavelson *et al.* [73] emphasized that self-concept is organized, or structured by domain, meaning that a person can have a different self-view depending on the context (e.g., physics class) and focus (e.g., computational activities). It is developmental, in that a person builds or develops a narrative about oneself in a particular set of contexts. Though it was at first used to describe broad self-views (i.e., self-esteem), self-concept was only later used to examine academic realms. Marsh and Craven [89] argued that what distinguishes academic self-concept is that students evaluate their performance in comparison to their performance in other domains, their peers' performances, and their internal standards of performance quality. Though focused on evaluation, it is distinguished from self-efficacy because the evaluation of performance is stabilized by previous evaluations and exists broadly for an entire school subject, whereas a self-efficacy judgment has more to do with prospective situations in a given academic domain. This would make the difference between self-concept and self-efficacy threefold: (i) domain-level versus task-level evaluation, (ii) evaluation of past performance versus prospective performance, and (iii) incorporation of evaluation into a sense of self versus a sense of ability.

In a theory-building paper by Brunner *et al.* [90], they propose and evaluate the effectiveness of a model for self-concept. The authors suggest using a first-order model (e.g., focusing broadly on academic self-concept) or a nested model (e.g., considering broad academic self-concept *and* math self-concept). They emphasize that self-concept can be split into separate self-concepts for each academic domain when using the nested model. In our context, this would indicate that this model of self-concept would be appropriate for the students who perceive computation as a separate domain from physics (not integrated *into* the domain of physics as a learning tool). This is in opposition to how Mr. Buford, the teacher, framed computation in his classroom.

While self-concept has not been used in computation research, there have been examples in other areas of education research. For instance, Chen and Xu [91] studied self-concept for junior high school English and its components: listening, speaking, reading, and writing. The qualitative case study of multiple students demonstrated how students with different self-concepts for different components can have drastically different trajectories in class, pointing to the complicated nature of self-concept for specific academic domains and activities. Espinosa [92] produced a quantitative study about cataloguing a variety of factors that build into academic STEM self-concept for college students. The core of her methods addressed self-concept from its most basic definition: evaluation of oneself. Mardiningrum [93] produced a case study on two participants in a university student theater club. The collaborative nature of this environment made social interaction a focal aspect of the participants' self-concepts. In a learning environment that uses group-based computation activities, we would expect social interaction to contribute to self-concept.

The studies above provide insight for how we might apply self-concept to a computation-integrated physics setting. The construct has not been used in this type of environment before, but we know that to apply it we need to focus on moments of self-evaluation [92], accounts of social interactions [93], and nuances in how students see themselves in relation to computational activities, computation, and physics as a whole [90,91]. This construct adds to our study because it can help us frame the way students discuss their feelings about computational experiences in a way that involves perceiving their role, as opposed to perceiving their ability (self-efficacy) or perceiving the malleability (or rigidity, in the case of fixed mindset) of their role and/or ability (mindset).

For example, in Sec. V C, Circe articulated that computation "doesn't make [her] feel like [she] belongs in physics." When students feel that they do not belong in a computation-integrated physics environment, they can also feel that they were not *meant* to belong there, as evidenced by Circe's later reflection on not having the brain for computation: "there's people who are really good at physics that are also really good at coding…I guess it's all the same kind of brain." This feeling is related to computation and/or physics self-concept because it could be framed as a perception of self in relation to a school subject. Feeling out of place in comparison to peers is part of self-concept [73]. The challenge lies in the potential for students to feel this way and lose interest in physics before gaining a realistic view of what it means to *do* physics.

Another challenge tied up with self-concept is interpreting code. Blaine lamented in Sec. V E about his feeling of inability to understand what the code meant. For Blaine, it was about feeling unable to make any progress on the activity and unable to interpret error messages. These roadblocks produced an affective response: Blaine said, "I don't know anything!" This evaluation of self in relation to computation indicates a self-concept judgment. Blaine felt stupid when doing computation.

Similarly, Blaine's made statements related to low self-concept in Sec. V D. Here, he outlined accumulation of negative experiences. Accumulations and patterns of experience are part of how a student builds self-concept for a school subject [73,89]. Blaine is a student who identified a pattern in his computational experiences: "I don't really try to do any more cause I'll put in a hundred things and then I'll just get a blank screen or I'll get some error." The repeated roadblocks with no success at overcoming them led Blaine to believe he "literally can't get anything but a blank screen." He suggested that he had experienced computation enough already to develop and hold this belief. Self-concept is tied to this challenge because the way Blaine's statements align with negative self-concept is tied to this pattern of experiences. It is important to acknowledge the ramifications when students deal with challenges unsuccessfully like this, one consequence being a development of self-views aligned with low self-concept.

As a final example, we look at Ed's delineation between physics and computation in Secs. V A and V B. In these sections, Ed said that the way she thought about physics "can't be programmed." This sends the message that not all physics knowledge is meant for a computer program; in particular, Ed's physics knowledge was not meant for a computer program. One possible theory-based explanation for this belief is that when Ed encountered a new method for representing and applying physics ideas (i.e., physics through computation), and when she found this new method to be uniquely difficult compared to her prior experience with physics, Ed protected her physics self-concept by building a separate, low self-concept for computational endeavors (or "GLOWSCRIPT," "coding," etc.). This separation can mean that some students do not let themselves develop as doers of computation, and it can prevent them from learning on days when this is an aspect of their physics class.

Self-concept suggests that students can develop a view of themselves in physics that is different from the view of themselves when doing computational activities, which validates the possibility of Ed's experience with separating her interpretation of the two domains. Because self-concept has not been applied to computation-integrated physics before, our work indicates it might be a viable lens for understanding how students are internalizing their experiences in computation. For example, future work could point to the process by which self-views related to self-concept develop in these settings, how students reconcile their views of the two different domains (physics and computation), and how that fits in with the theory of broader academic self-concept.

### D. Intersection of self-efficacy, mindset, and self-concept

In talking about the challenges that they faced, the students in our data made statements that point to their views related to the theories of self-efficacy, mindset, and self-concept. While we previously discussed these constructs as separate ideas, we want to emphasize that these are not independent theories or constructs. In fact, the overlap between these constructs illuminates avenues for future research, curriculum design, and pedagogy.

For example, we can see aspects of all three constructs in how Blaine faces the repeated confusion challenge. Blaine described how he experienced a series of failures related to doing computation: "I literally can't get anything but a blank screen…I'll put in a hundred things and then I'll just get a blank screen or I'll get some error." These failures fit narratives about the reduction of self-efficacy and negative impact on self-concept, and the way Blaine articulates them aligns with the language of fixed mindset. Each framing provides a different insight into Blaine's description of his experience. The self-efficacy framing shows the impact of serial mastery experiences on views related to self-efficacy, as shown when Blaine described how he felt that he "literally can't get anything but a blank screen" after repeatedly failing to make progress in the computational activity. The self-concept framing shows how a pattern of negative experiences can come to define what computation means to a student in the moment, as shown when Blaine expressed apathy when describing his relationship with computation: "I just don't even care. I'm like 'whatever dude. I can't do this sh*t.'" The mindset framing brings focus to the parts of Blaine's behavior related to aspects of mindset, specifically the reduction of effort in response to his failures, which relates to fixed mindset: "I don't really try to do any more cause I'll put in a hundred things and then I'll just get a blank screen or I'll get some error." From this one example, we can see that the three frameworks overlap and build into one another. Blaine's repeated failures to make progress with the computation led to a reduction of effort and no other change in strategy, aligning with aspects of fixed mindset. This accumulation of failures also ties to his statements about his work and of himself—statements which align with having a lack of self-efficacy and/or a low self-concept for computation.

This illustrates how the theoretical lenses can overlap and provide a fuller picture of the impact that the affect-based challenges can have on students. We use all three to highlight different views on the same individual experiences, but they provide varied angles from which to understand what is going on with the students in our study. That said, this study only provides an initial window into how these frameworks relate to one another, and we suggest future research specifically focus on how each framework fits with one another in this context, how theory-based interventions might impact students' perceptions, and how these frameworks might be leveraged to better understand computation-integrated classrooms. We view the presence of many angles as a way to identify jumping-off points for further research on affect-based learning and challenges, which is sorely needed and which we highlight in the discussion section. However, we first highlight some positive experiences that students recounted in their interviews. These did not fit in with our challenges, but still provide a unique perspective on what students experience and how computation can be beneficial, according to students.

## VII. POSITIVE STUDENT EXPERIENCES

Along with the challenges students faced and recalled in their interviews, there were also indications of positive experiences brought on by computation. In this section, we outline a handful of beneficial impacts of computational integration that students interpreted. Afterwards, we discuss how they relate to some of the goals that Mr. Buford set out to achieve by introducing computational activities to his class.

We begin with a comment from Ed that demonstrates how she learned about using computation to see physics. She describes getting "the whole concept of coding" through engaging in a computational activity about collision physics.

> Ed: We were doing momentum, and we were looking at elastic and elastic collisions, and we actually coded something where two blocks had to collide. And just seeing how just changing a couple of numbers could change the entirety of the coding was interesting… That was helpful for me to get the whole concept of coding.

> Interviewer: What concept did it help you understand better?

> Ed: Momentum itself, its maximum velocity. I don't know. Seeing the effects typing in different numbers had on each block, and making it go faster, or making it go slower, changing the masses of the box. It just helped me with the concept of it.

Ed came to understand how changing numbers in the program is connected to seeing the physical consequence in the animation. Computation allowed her to make small changes to the program and to see the relationship between momentum and the actual movement of objects. Ed's articulation of this and engagement at this level suggests an orientation towards learning physics through computation rather than just trying to get through the activity. Though she outlined many challenges in the previous section, this comment shows that students also see benefits to computation, and one of those benefits is the visualization and strengthening of physics concepts.

Joyce expressed a similar perspective, which was that the process of translating ideas into code was a way of learning physics concepts. While Ed focused on the benefits of interacting with the dynamic, completed code, Joyce discussed how creating code was constructive for her.

> Joyce: By actually coding the formula and what variables go in, I think it helps in learning the concepts. It's just you might not catch [an error] at first and you might mess up because we were supposed to put other stuff in [the program].

Joyce shared how she felt like she learned the physics concepts better by coding the formulas and variables. In the second part of her quote, Joyce talked about the experience of accidentally letting a bug, or coding error, get into the program ("we were supposed to put other stuff in the program") and prevent it from running properly. By relating learning physics concepts to the debugging process, Joyce demonstrated that she understood there was value in meticulously translating physics formulas into code and incorporating the computer's feedback. This awareness allowed her to engage with the activities in a way where she felt that they helped her learn physics.

Finally, we found computation can help some students build interest in physics. Beck discussed at length how he viewed computation as an opportunity to connect with physics in a more authentic way. Below, he talked about how a visual world of physics opened up when he used GLOWSCRIPT.

> Beck: GLOWSCRIPT provided even more visuals and stuff to actually connect with, which is what made me understand physics and like it even better. The visuals, the demonstrations, that ability to see the things in real life… they just helped provide even more for that, and they even strengthened my liking for physics even more.

He connected with the visuals and felt as though he was seeing the phenomenon in real life. Beck went on to say more about the benefit of computation, describing how it provided an opportunity to do some of the same activities that physicists do professionally.

> Beck: [Coding] allows you to apply stuff that you've learned in a way that's different from just solving a problem on paper, because you actually get to see the result of what you've solved in real life. I mean it's a computer, but you get to see it actually work. It gives you a view of what physicists do, I suppose. Like you get a problem and you use physics to solve the problem, then you see it actually work… I like the coding in physics because of that.

In this excerpt, Beck saw the purpose of computation as seeing a physics problem at work in a simulation of the real world. It was a way for him to connect what he was learning to what was relevant to him. It was also a way to understand the type of work that actual physicists do. In Beck's case, this engendered an interest in him saying "I like the coding in physics" and "strengthened my liking for physics even more." This shows that computation has the potential to help students build an interest in authentic physics as well as help with learning.

The benefits that Ed, Joyce, and Beck described are similar to some of the goals that Mr. Buford had for his computational integration. In particular, he wanted students to strengthen their understanding of physics concepts through computation, saying, "I hope it just enhances them thinking about the physics concept that we're trying to learn, ideally….I feel like when you're writing the code for this, you have to understand how projectile motion works, or you can't write code that models that very well." Both Ed and Joyce described the benefit to conceptual understanding, though it is not clear whether Mr. Buford envisioned the same mechanisms of learning. For Ed, she learned through interacting with the completed code, and for Joyce, learning happened through creating the code itself and working through bugs. The benefit that Beck described goes beyond what Mr. Buford said, namely the computation helps him do *real* physics and builds his interest in the subject.

There were also some ideas that were missing from student interviews, benefits that Mr. Buford envisioned but that did not seem to bear out in our data. Mr. Buford hoped that the open-ended nature of the computational activities and the choice to not grade them would spur students to be more creative, given that a lot of the constraints on traditional physics projects were stripped away. Students did not seem to latch onto the creative freedom in their interviews, so it is unclear to what degree this goal was realized in the actual implementation. Also, there remains the question of what benefits could exist in other implementations. For example, Mr. Buford wondered whether "you could use [computation] as a way of developing concepts" rather than just reinforcing. With different design goals and in different contexts, this might be entirely possible, which could chain into students seeing different benefits to the computational integration.

## VIII. DISCUSSION

From students' interviews, we see that they faced a variety of challenges when computation was integrated into their physics class. Some of these challenges were related specifically to code (e.g., interpreting code, repeated confusion), while others were related to the pedagogy and culture of the classroom (e.g., interpretations of implementation), but many of them were unique or had unique components due to the integrated physics classroom context (e.g., feeling worse at physics, unbelonging and stereotypes, stress or frustration).

The challenges that we found specifically related to code (interpreting code and repeated confusion) are similar to the student challenges reported from computer science contexts. Jenkins [36] highlighted barriers in introductory level computer science learning, mainly focusing on the extra skills that students need to learn to engage with computation, such as syntax, semantics, and algorithms. He argued that what made computation hard was chiefly the novelty of it. This aligns with what we found in Mr. Buford's computation-integrated physics class. For example, a part of interpreting code is understanding syntax and how it pieces together as well as error messages and strategies for addressing them. These are new skills that students did not encounter before unless they took a computer science class. Even then, we found that students who *had* taken a computer science course still struggled with the syntax and idiosyncrasies of GLOWSCRIPT. Previous research by Bumler *et al.* [94] found that students with prior computational experiences did not view minimally working programs using the GLOWSCRIPT platform as authentic computation. The conflict between their previous experiences and the lack of utility of students' previous experiences in the context of this research implies there are difficulties transferring practice to the GLOWSCRIPT platform. The basis for this disconnect between platforms and contexts needs to be studied in greater detail. This also speaks to the repeated confusion challenge because the process of learning a programming language (especially debugging) requires persisting through many mistakes and learning from them. This parallels another study, in which Bosse and Gerosa [29] cataloged some of the main worries that students tend to have in programming settings, including trouble with syntax, variables, error messages, and code comprehension. The worries were sometimes so overwhelming that when a student realized their code contained an error, they were more likely to give up. We saw a similar case with Blaine, who gave up after encountering numerous errors and no longer proceeded with the activity. From another perspective, Svensson *et al.* [48] viewed computation as a social semiotic, or a way of communicating about and exploring phenomena. They saw challenges emerge when students had limited skill with using the semiotic resources, even when students *did* see the benefit of communicating and exploring through computation. This mirrors the experiences of Ed and Otto, who both saw the usefulness of computation and often even knew the relevant physics concepts, but they ran into roadblocks because they had limited experience and comfort with computation itself and/or GLOWSCRIPT. The fact that we saw the same challenges and barriers in the computation-integrated environment that are seen in computer science contexts indicates that students' interpretation of code is a broader challenge for any type of coding activity. Given the common challenge between contexts, this would indicate a place where computer science educators and physics educators can learn from one another about how to best support students.

However, we also found several challenges that were unique to the computation-integrated physics environment. For example, in Sec. V B, Ed separated the domains of computation and physics, so that her difficulty with computation would not affect her view of her physics competence. She reassured herself, "You know [the physics], you just think about it in a different way, but that's not a way that can be programmed on the platform." This challenge was unique to the computation-integrated physics environment, specifically because the curriculum merges two subjects that for these students can sometimes be viewed as two separate domains. In a separate computer science course, students' perceived physics competencies and self-views would typically not be threatened or involved at all. However, because of the integration, some students protected one view of themselves, potentially at the cost of the other. In the case of Circe, the integration of computation led to statements of unbelonging and a distancing of herself from physics as a whole. In the case of Blaine, we saw that multiple failures at the computational activities led to statements aligned with lack of self-efficacy and low self-concept when he said, "I don't know anything!" This is similar to what Lishinski *et al.* [30] found during computational activities (and what Caballero *et al.* [7] warned against), namely, that judgments aligned with lack of self-efficacy can lead students to use tactics that harm their learning rather than help.

The integration of computation into STEM is strongly motivated, including arguments about preparation for students' future careers and making STEM courses more relevant. However, we showed that students intentionally separated the domains at times. This would indicate to teachers, researchers, and curriculum developers that more attention needs to be directed to *how* this integration occurs. For example, as part of the ICSAM workshop, Mr. Buford was altering an existing curriculum. He already had lesson plans to teach all the necessary physics content, and perhaps it made more sense to introduce computation at the transition points in the curriculum rather than potentially disrupt the material midconcept. Additionally, ICSAM teachers learned how to program with GLOWSCRIPT during a summer workshop. They were already physics experts when they arrived, but many were novices at computation, meaning they learned to program as a way of modelling and exploring what they already knew about physics. This process could have transferred to how their students would go on to learn computation in their classrooms: physics first, computation later. Ultimately this could have contributed to the separation of computation and physics as separate domains. That said, there is certainly a precedent for integrating STEM domains. After all, physics and math have been closely tied since the foundation of the field. We do not think twice about whether formulas and calculations are a part of physics, and for students, learning to use math as a tool and learning physics go hand in hand. In the same way, we envision a future where computation is also treated as an everyday tool for learning physics in classrooms and viewed as such by students, but we need to learn more about what is happening in these integrated classrooms. However, the math and science domains are blended at a much earlier point in a student's schooling. Students perceiving computation and physics as two different domains highlights the need to investigate whether integrating at an earlier point in a student's science careers would impact their perceptions of computation being a tool for doing science.

Another challenge that is unique to the computation-integrated contexts is the balancing of content between computation and physics. Given the other constraints that teachers are under (time limitations, science standards that must be met, etc.), it can be difficult to add computation to an already packed schedule. Mr. Buford commented in his interview that despite his natural curiosity for new ideas in physics, it was hard to try new things when he had to cover all the content on the AP Test. The year before he attended ICSAM, he simply saved computation until after the test was over in the last month of the school year. When he tried to integrate computation into his curriculum throughout the year, he was not able to let the students slow down enough to wrestle with the computation and figure out how it could help them learn physics. For some students, the purpose of doing computation in physics did not stick with such little time. Furthermore, there might be some influence from the AP curriculum on what counts as "doing physics." Without changing the national expectations and standards to include computation, it will be near impossible to create fully integrated courses.

We also saw several challenges that were related to pedagogical choices from Mr. Buford. For example, Mr. Buford intentionally chose to not grade the computational activities, which led to Ed commenting that the activities felt like "busy work." He also chose to show the final output to the class, and when Circe's answers did not match, this made her feel like her answers were "wrong." In Sec. V A we highlighted how the students felt as though the activities being after the concept had been covered had framed them as causing undue stress. However, from Mr. Buford's perspective this was intentional because he thought introducing concepts via a computation activity would be too stressful. This "catch-22-like" outcome highlights the struggle that teachers face when making curriculum design decisions around integrating computation into their classrooms and highlights a desperate need for research focused on curriculum design for such environments. None of the above discussion points around pedagogical choices is intended as a critique of Mr. Buford (in fact he had strong pedagogical reasoning for his choices), but this highlights that there might be unique challenges depending on the specific implementation of computation-integrated physics and the classroom structures that a teacher employs.

For example, Beck described a positive structure in his interview from Mr. Buford's class. Beck came upon a roadblock and had to ask for help from Mr. Buford, who pointed out to him a built-in GLOWSCRIPT function that did *exactly* what he needed. In fact, having students ask for this function was part of Mr. Buford's plan—he confirmed after class to the first author that part of the activity's purpose was to discover the need for a new function. The challenge lies not in what we generated in the data, but in what was absent: the students who did not think to ask for help or who did not arrive at the point in the activity to realize the need for a special function. Students might struggle to ask for help for a variety of reasons; they might feel intimidated by asking questions to an authority figure (their teacher in this case), they might feel too embarrassed by their "lack of progress" on the problem to ask for help, or they might struggle with social anxiety. Alternatively, and especially in a less collaborative context, students might have the impression from classroom norms or social stereotypes that they are supposed to be coding alone. Any of these reasons might prevent students from asking for help, and in turn, increase their frustration and perpetuate a negative view of computation.

As Mr. Buford confirmed, a teacher might let students struggle with an idea intentionally or might want students to discover an idea as part of the computational activity. With Beck, this worked well, and he was able to learn about the unit vector from Mr. Buford. However, for this to happen it was critical that Beck felt comfortable asking Mr. Buford for help and that Mr. Buford promoted that in his classroom. In another classroom context, with a different classroom culture, we could envision "asking for help" to be a challenge for students. This only points to the work that needs to be done to build on this study and examine the contextual challenges in other implementations of computation-integrated physics and other STEM courses.

## IX. CONCLUSIONS AND FUTURE WORK

In this paper, we have described the student-perceived, affect-based challenges that high schoolers faced in a computation-integrated physics class: Stress or frustration, feeling worse at physics, unbelonging and stereotypes, repeated confusion, interpreting code, and interpretations of implementation. We also found connections between students' descriptions of the challenges and the theories of self-efficacy, mindset, and self-concept. This work is laying the foundation for identifying affective barriers and those unique to computation-integrated STEM contexts, serving as the first study in this context to examine affective challenges from students' perspectives. While this study is an initial step, more work needs to be done to understand the affective challenges students face and how to best support them.

An example of the importance of student perspectives from our data was when Joyce said she felt "just average" at coding when we were fleshing out the unbelonging and stereotypes challenge. She appeared to be one of the most competent programmers in class, but she did not feel that way about herself. It is only through asking students about their experiences that we can find out how they feel about the challenges they face in class, and sometimes their answers can be unexpected.

In this study, students' perspectives demonstrated that for some students in an iteration of classroom computation, challenges can overshadow positive experiences. These challenges deserve to be addressed as curriculum developers continue to integrate computation into physics courses. Though the context presented here did not span beyond high school physics, we know that computational integration spans K–16 in disciplines from physics and biology to language arts and social studies [1–4,7,10,12,95,96]. Such integration is bringing modern ways of understanding and applying disciplinary concepts to the classroom. We believe such curricular transformations are necessary. In some cases, these transformations can make disciplinary content relevant, nurture new curiosity, and lead to new modes of understanding, as shown in Sec. VII. The question at this point is, how do we move forward and structure computational opportunities to be helpful, positive, equitable, and meaningful?

To researchers, this study is a call to action. Computation-integrated physics courses continue to grow as computation becomes synonymous with doing STEM. With it come the complexities and difficulties of new curriculum and the need to understand the experiences of students in this new environment. We have found that the lenses of mindset, self-efficacy, and self-concept might offer meaningful insight into many student-centered processes, yet there is a need for more exploration, particularly in how the integration takes form, how the protective separation of computation from physics can be minimized, and how the difficulties and frustrations of learning a new programming tool affect students. We need studies on affect, self-beliefs, and perceptions in computation-integrated contexts where computational learning is supported by design, where the curriculum is less constrained institutionally (e.g., "regular" instead of AP), where computational tools are the focus of the course, and where features of implementation support underrepresented students.

To practitioners, this study is a call to consider many factors when designing or altering curriculum for computational integration. We call for attending to the affect of students who take part in the curriculum, the tools being used to integrate computation, the pedagogical strategies for teaching computation, what it means to redesign existing curriculum, the curriculum's potential effect on students' perceptions of computation and physics, and the role computation can play pedagogically. We acknowledge that figuring out how computation best fits into the context of one's physics course is an immense task. We need to teach students authentic physics by using computational tools, but we also need to find ways to ease the burden on physics teachers who are often saddled with altering curriculum to meet new educational demands, of which computational integration is the latest [47].

In conclusion, we highlight that the computational challenges raised in this paper need to be studied in more depth in the computation-integrated context as opposed to trying to understand them by only applying knowledge from physics or computer science education research. This type of curriculum is unique enough to warrant further studies, especially when considering the issues that arose when students had to deal with computation and physics at the same time in the same context. Computation in our physics courses is essential for the next generation of scientists, and it is imperative that we learn how to best apply computation as an educational tool to the benefit of our students.

## ACKNOWLEDGMENTS

We acknowledge and give thanks to the National Science Foundation (DRL-1741575) for funding this research, to Mr. Buford (pseudonym) for providing access to his classroom and eagerly participating in the research, and to the students who gave their time and insights to us during the interview process. We also thank Jacqueline Bumler for assisting with data generation and providing additional insight during discussions about the data. Lastly, we thank David Stroupe, Vashti Sawtelle, and Tyce DeYoung for feedback on the manuscript.

## APPENDIX

This is the original interview protocol used to conduct semi-structured interviews with the student participants.

1. Tell me about yourself.
   (a) What year are you in school?
   (b) Why did you choose to take this physics course?
   (c) Have you taken a physics course before this one?
   (d) What do you want to do after high school?
2. Tell me about what you do in physics class.
   (a) Are there different sorts of activities you do? Can you describe them for me?
      i. Do you always solve for a number? Do you have to design things?
      ii. Do you ever work with equipment?
      iii. Do you always work by yourself, or do you work with your classmates?
      iv How do you interact with your classmates?
      v How do you interact with the instructor?
   (b) How is this class different from prior physics classes?
   (c) Do you think you're good at physics?
   (d) Are there times you struggle more than others?
   (e) Are there things you do in class that make you feel as if you can or can't do physics?
   (f) Are there times in class when you feel more like a scientist/physicist?
3. About the computational activities in Mr. Buford's class…
   (a) Why do you think Mr. Buford added computational activities to the class?
   (b) Have you done anything with computation before?
   (c) Was there anything new or exciting that you were able to do with computation? Can you give an example?
   (d) Do you like the computational activities? Why or why not?
   (e) Do you ever get frustrated in class? What has frustrated you and why?
   (f) Do you think you're good at computation?
   (g) Are there things you do in class that make you feel as if you can or can't do computation?
4. When you get stuck with computation, what do you do?
   (a) Do you wait until Mr. Buford can help?
   (b) Do you try to consult with your group mates?
5. What do you learn/gain during the coding days?
   (a) What about regular days?
   (b) What do you learn? How do they differ?
6. How do you tend to participate in class?
   (a) When Mr. Buford is talking to the class?
   (b) When you are working together in a small group?
   (c) Does this change when you are doing computational activities as a group?
      i. What role do you take on when the group doing computation?
7. What if you were told that computation is a big part of what you want to do in the future?
8. What is a subject you really like (or really don't) and how does your experience in that class compare to physics class?
9. Have you done computation before Mr. Buford's class? How did you feel about it?